\documentclass[twocolumn,preprintnumbers,amsmath,amssymb,floatfix]{revtex4}

\usepackage{graphicx}
\usepackage{dcolumn}
\usepackage{bm}
\usepackage{textcomp}
\usepackage{booktabs}

\begin{document}

\title{
Phase Diagrams and Ordering in Charged Membranes:\\
Binary Mixtures of Charged and Neutral Lipids\\
}
\author{Naofumi Shimokawa\footnote{nshimo@jaist.ac.jp}}
\author{Hiroki Himeno}
\altaffiliation[Also at ]{Health Research Institute,
National Institute of Advanced Industrial
Science and Technology (AIST),
Kagawa 761-0395, Japan}
\author{Tsutomu Hamada}
\author{Masahiro Takagi}
\affiliation{%
School of Materials Science, Japan Advanced Institute of Science and Technology,
Ishikawa 923-1292, Japan
}%
\author{Shigeyuki Komura}
\affiliation{%
Department of Chemistry,
Graduate School of Science and Engineering,
Tokyo Metropolitan University, Tokyo 192-0397, Japan
}%
\author{David Andelman}
\affiliation{%
School of Physics and Astronomy, Raymond and Beverly Sackler Faculty of Exact Sciences,
Tel Aviv University, Ramat Aviv 69978, Tel Aviv, Israel}%

\date{\today}
\date{May 24, 2016}

\begin{abstract}
We propose a model describing the phase behavior of two-component membranes
consisting of binary mixtures of electrically charged and neutral lipids.
We take into account the structural phase transition (main-transition) 
of the hydrocarbon chains, and investigate the interplay between this 
phase transition and the lateral phase separation.
The presence of charged lipids significantly affects the phase behavior
of the multicomponent membrane.
Due to the conservation of lipid molecular volume, the main-transition temperature
of charged lipids is lower than that of neutral ones.
Furthermore, as compared with binary mixtures of neutral lipids, the membrane
phase separation in binary mixtures of charged lipids is suppressed, in accord with
recent experiments.
We distinguish between two types of charged membranes: mixtures of charged saturated
lipid/neutral unsaturated lipid and a second case of mixtures of neutral saturated lipid/charged unsaturated lipid.
The corresponding phase behavior is calculated and shown to be very different.
Finally, we discuss the effect of added salt on the phase separation and the temperature
dependence of the lipid molecular area.
\end{abstract}

\maketitle

\section{Introduction}
\label{intro}

Biomembranes are composed of various components such as
phospholipids, sterols, and proteins, and are believed to have
compositional heterogeneities.
In multicomponent membranes, regions or domains enriched in
saturated lipid and cholesterol are called ``rafts", and are thought
to play an important role in various biocellular functions,
such as signal transduction and membrane trafficking~\cite{SI,SS}.
Artificial bilayer membranes composed of several phospholipids and
cholesterol have been used as typical model systems for biomembranes.
Phase separation in such model membranes has been widely
investigated in order to understand the mechanism of raft
formation~\cite{Komura14}.
Considerable effort has been given to ternary systems composed of
saturated lipid, unsaturated lipid and cholesterol, where it is
known that domains enriched in saturated lipid and cholesterol
form a liquid-ordered phase in a matrix of an otherwise
liquid-disordered phase~\cite{Keller1,Keller2,Baumgart,HF,Hamada11a,Hamada11b,Shimokawa15}.

Two-component lipid bilayers consisting of saturated and unsaturated
lipids also show a phase separation between solid-like (L$_\beta$) and
liquid-like (L$_\alpha$) phases.
This phase separation can be directly visualized by fluorescence
microscopy, which demonstrates that the solid-like L$_\beta$ domains
exhibit various anisotropic shapes~\cite{BG,Poon,CS1,CS2}.
The resulting phase diagrams of different binary mixtures have been
experimentally explored in great detail~\cite{CS2,WM,LM,ALG,BWGG},
and have been quantitatively reproduced in several theoretical
works~\cite{KSOA,KSO,ESS,SKA1,YBS}.
In particular, in models proposed by some of the present authors~\cite{KSOA,KSO,SKA1},
a coupling between the membrane composition and its internal structure
was suggested in order to consider the interplay between the phase
separation and the chain phase transition (main-transition).
In these experimental and theoretical studies, however, the
lipids were taken to be electrically neutral.

Since a certain amount of charged lipid is always present in
organelles such as mitochondria and lysosomes~\cite{AlbertsBook},
electrostatic interactions are significant in biomembranes.
Recently, several experiments have investigated the phase separation
of multicomponent lipid mixtures containing at least one type of
charged phospholipid~\cite{Dimova1,SHSY,Keller4,Dimova2,Himeno,Himeno15}.
In binary mixtures of neutral saturated lipid and negatively charged
unsaturated lipid, the phase-separated region was suppressed as
compared with the case when both lipids are neutral.
This phase-separated region reappears when salt (e.g.,\ NaCl),
which screens the electrostatic interactions, was added.
This is a strong indication that the electrostatic interaction affects
the phase behavior.
Although such behavior has been qualitatively explained by
phenomenological models using the Poisson-Boltzmann (PB)
theory~\cite{SHSY,May,SKA2}, the structural phase transition
(the main-transition) of the membrane was not considered.

\begin{figure}[tbh]
\begin{center}
\includegraphics[scale=0.45]{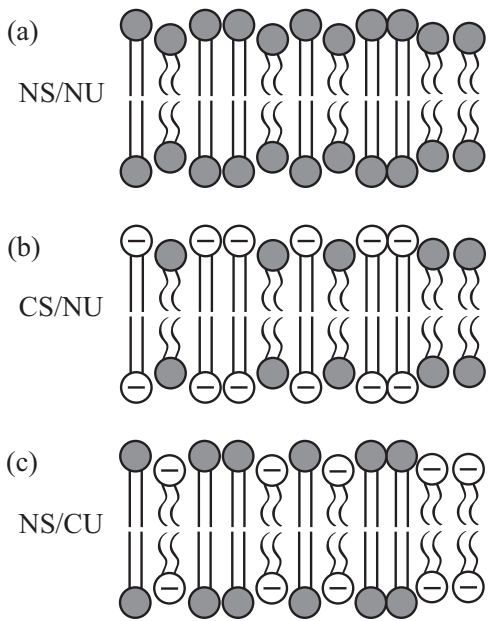}
\end{center}
\caption{Schematic illustration of two-component membranes.
(a) A neutral membrane consisting of a binary mixture of neutral saturated
(NS)  and neutral unsaturated  (NU) lipids.
(b) A charged membrane consisting of a binary mixture of charged saturated
(CS)  and neutral unsaturated  (NU) lipids.
(c) A charged membrane consisting of a binary mixture of neutral saturated
(NS)  and charged unsaturated  (CU) lipids.
}
\label{fig1}
\end{figure}

In membranes composed of binary mixtures of neutral and charged lipids,
either the saturated or unsaturated lipids can be charged.
Hence, it is convenient to define two types of charged membranes:
(i) mixtures of charged saturated (CS) and neutral unsaturated (NU) lipids
(denoted as CS/NU, see Fig.~\ref{fig1}(b)), and
(ii) mixtures of neutral saturated (NS) and charged unsaturated (CU) lipids
(denoted as NS/CU, see Fig.~\ref{fig1}(c)).
In recent experiments by Himeno \textit{et al.}~\cite{Himeno}, it was reported
that the phase separation was suppressed in the NS/CU lipid mixture as compared
with the neutral membrane (NS/NU as in Fig.~\ref{fig1}(a)), while it was enhanced
in the CS/NU case.
This result indicates that the phase behavior of charged membranes is
very different between these two types of charged mixtures.

Being motivated by this experimental work, we consider in this paper charged
membranes, and propose a phenomenological model that distinguishes between
the above-mentioned two cases.
In particular, we discuss the interplay between the main-transition and the lateral
phase separation between the two lipid species.
The main-transition of lipid chains is taken into account by extending the
model of Goldstein and Leibler~\cite{GL1,GL2} to include electrostatic
interactions using the PB theory~\cite{DA1,DA2}.
Furthermore, we employ the constraint of constant molecular volume of the
lipids, which leads to a coupling between the electrostatic interactions
and the lipid lateral phase separation.
We calculate the different phase diagrams of charged membranes by changing the
interaction between the two lipids and/or by changing the amount of added salt.
We also explain the different combination of hydrocarbon saturation and headgroup
charge (either CS/NU or NS/CU as were defined above).
On the basis of the calculated phase diagrams, we discuss the main-transition temperature,
and the effect of added salt in a binary mixture of charged/neutral lipids.

In the next section, we explain the model by Goldstein and Leibler, the phase
separation, and the PB theory.
In Sec.~\ref{results}, we show the calculated phase diagrams and
investigate the effects of the interaction between the two lipids,
salt concentration, and combination of the lipids (either CS/NU or
NS/CU) on the phase separation.
Moreover, we explore how the molecular area changes and its dependence on
the added salt.
Finally, we offer a qualitative discussion in Sec.~\ref{discussion} and some concluding remarks in
Sec.~\ref{conclusion}.

\section{Model}
\label{model}

In order to calculate the phase diagrams of binary mixtures of charged
lipid membranes, we consider a model that takes into account both the
lipid main-transition and the lateral lipid phase separation.
For simplicity, we consider a perfectly flat membrane ignoring any possible
out-of-plane membrane undulations.
To avoid further complexity, we do not consider any specific bilayer
attributes such as asymmetric lipid composition between the two
leaflets, or ionic strength difference between the electrolytes in
contact with the two sides of the membrane.
Therefore, our model describes the phase behavior of an isolated
symmetric bilayer.

\subsection{Two-Component Neutral Membranes}

On the basis of the Goldstein-Leibler model~\cite{GL1,GL2}, we first review
the free energy of a single-component neutral lipid membrane, which leads
to the main-transition.
The main-transition involving the chain ordering and stiffening is described
via a scalar order parameter
\begin{equation}
\label{op}
m=\frac{\delta-\delta_{0}}{\delta_{0}},
\end{equation}
where $\delta$ is the actual membrane thickness and $\delta_{0}$ a reference
membrane thickness in the L$_{\alpha}$ phase.
Notice that $m$ represents the changes that may occur in several degrees
of freedom,  including the conformation of the hydrocarbon chains, their
interchain correlations, molecular tilt, and positional ordering.
For an isolated lipid bilayer, a Landau expansion of the stretching
freeenergy per lipid molecule in powers of the order parameter $m$ is given by
\begin{equation}
\label{stretch}
f_{\rm st}(m)=\frac{a'_{2}}{2}(T-T^{\ast})m^{2}+\frac{a_{3}}{3}m^{3}
+\frac{a_{4}}{4}m^{4},
\end{equation}
where $a'_{2}>0$, $a_{3}<0$, and $a_{4}>0$ are phenomenological
coefficients.
Here $a_{3}$ is taken to be negative since the main-transition is a
first-order transition~\cite{LM}, while $a_{4}$ is always positive to
ensure thermodynamical stability.
In the above, $T$ is the temperature and $T^{\ast}$ is a
reference temperature.
The main-transition temperature is different from $T^{\ast}$ and is denoted
by $T^{\ast\ast}$.
It can be calculated from the conditions $f_{\rm st}=0$ and ${\rm d}f_{\rm st}/{\rm d}m=0$,
yielding
\begin{equation}
\label{tm}
T^{\ast\ast}=T^{\ast}+\frac{2 (a_{3})^{2}}{9a'_{2}a_{4}}.
\end{equation}
Note that the main-transition temperature described by Eq.~(\ref{tm})
is appropriate only for neutral lipids.

Next, we describe the phase separation in membranes containing a binary
mixture of neutral saturated  (NS)  and neutral unsaturated (NU) lipids,
as shown in Fig.~\ref{fig1}(a).
We model the two-component membrane as an incompressible NS/NU lipid mixture
containing $\phi$ molar fraction of NS lipid and $1- \phi$
molar fraction of NU lipid.
(Throughout this paper, $\phi$ denotes the molar fraction of saturated lipid.)
For simplicity, we assume the same area per molecule for
the two species in the L$_{\alpha}$ phase and ignore any lipid exchange with the
surrounding solvent.
The mixing free energy (per lipid molecule) consists of the mixing entropy and enthalpy.
It can be expressed within mean-field theory as
\begin{equation}
\label{bw}
f_{\rm mix}(\phi)=k_{\rm B}T[\phi \ln \phi +(1- \phi) \ln (1- \phi)] +\frac{J}{2} \phi (1- \phi),
\end{equation}
where $k_{\rm B}$ is the Boltzmann constant and $J>0$ is an
attractive interaction parameter between the lipids, which
enhances the NS/NU demixing.

In general, the two lipids have different main-transition temperatures originating
from different molecular parameters such as their chain length, degree of saturation and
hydrophilic headgroup.
The dependence of the main-transition temperature on the
composition $\phi$, cannot be
calculated from such a phenomenological approach.
Therefore, we proceed by further assuming the simplest linear interpolation
of $T^{\ast}$ as a function of $\phi$:
\begin{equation}
\label{linear}
T^{\ast}(\phi) = \phi T_{\rm S}^{\ast} + (1-\phi) T_{\rm U}^{\ast},
\end{equation}
where $T_{\rm S}^{\ast}$ and $T_{\rm U}^{\ast}$ are the
reference temperatures of pure NS and NU lipids, respectively.
In general, saturated lipids have higher transition temperatures than the
unsaturated ones, $T_{\rm S}^{\ast\ast}>T_{\rm U}^{\ast\ast}$, for the
cases when the numbers of hydrocarbons in the lipid tails are the same.
Therefore we generally use $T_{\rm S}^{\ast}>T_{\rm U}^{\ast}$.

By combining Eqs.~(\ref{stretch}) and (\ref{bw}), we obtain the total free energy
of a neutral NS/NU membrane
\begin{equation}
f_{\rm NS/NU}(m,\phi)=f_{\rm st}(m,\phi)+f_{\rm mix}(\phi).
\label{neutral-total}
\end{equation}
This total free energy will be minimized with respect to both $m$ and $\phi$ to 
obtain the phase behavior of a neutral membrane. 

\begin{figure}[bth]
\begin{center}
\includegraphics[scale=0.45]{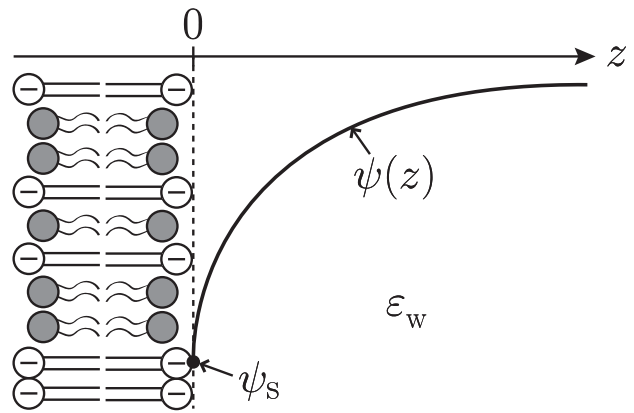}
\end{center}
\caption{Schematic representation of the electrostatic potential of
a two-component charged membrane consisting of charged saturated (CS)
and neutral unsaturated (NU) lipids.
The lipid head groups are located at the plane $z=0$.
The dimensionless electrostatic potential in solution is $\psi(z)=e\Psi/k_{\rm B}T$,
and $\psi_{\rm s}=\psi(z=0)$ is the surface potential at the membrane.
The dielectric constant of water solution is $\varepsilon_{\rm w}$.
}
\label{fig2}
\end{figure}

\subsection{Electrostatic Free Energy of a Charged Membrane}

In what follows, we extend the above model for the case of membranes consisting of
a binary mixture of charged and neutral lipids.
In order to estimate the electrostatic interactions in membranes at a fixed surface charge
density (chosen to be negative throughout this paper), $\sigma<0$, we obtain the
free energy of a charged membrane on the basis of the PB theory~\cite{DA1,DA2}.

As shown in Fig.~\ref{fig2}, we consider a flat charged lipid bilayer that lies at $z=0$.
The bilayer is in contact with a monovalent salt solution located at $z > 0$.
We assume that the two leaflets constituting the bilayer do not interact with
each other.
Within a mean-field treatment, the electric potential $\Psi(z)$
satisfies the PB equation
\begin{equation}
\label{PB}
\frac{{\rm d}^{2} \Psi}{{\rm d} z^{2}}=\frac{2en_{\rm b}}{\varepsilon_{\rm w}} \sinh \frac{e \Psi}{k_{\rm B}T},
\end{equation}
where $e$ is the elementary charge, $n_{\rm b}$ the salt concentration in the bulk, and
$\varepsilon_{\rm w}$ the dielectric constant of water.
Using the dimensionless electric potential
$\psi = e\Psi/k_{\rm B}T$, we obtain the following
dimensionless PB equation
\begin{equation}
\label{dimensionless}
\frac{{\rm d}^{2} \psi}{{\rm d} z^{2}}=\kappa^{2} \sinh \psi,
\end{equation}
where $\kappa^{-1}=\sqrt{\varepsilon_{\rm w}k_{\rm B}T/2e^{2}n_{\rm b}}$
is the Debye screening length.
Far from the membrane, the bulk electric potential should vanish,
i.e., $\psi(z \rightarrow \infty)= 0$.
With the use of the Gauss law, the boundary condition at the
membrane ($z=0$) is
\begin{equation}
\label{gauss}
\left. \varepsilon_{\rm w} \frac{{\rm d}\psi}{{\rm d}z} \right|_{z=0}=
-\frac{e\sigma}{k_{\rm B}T}.
\end{equation}
Under these boundary conditions, the dimensionless electric potential $\psi(z)$
can be obtained~\cite{DA1,DA2,Colloid} by integrating twice Eq.~(\ref{dimensionless})
\begin{equation}
\label{electric_potential}
\psi(z) = 2 \ln  \frac{1+\Gamma \exp(-\kappa z)}{1-\Gamma \exp(-\kappa z)},
\end{equation}
where
\begin{equation}
\Gamma = \frac{\exp(\psi_{\rm s}/2)-1}{\exp(\psi_{\rm s}/2)+1}.
\end{equation}
Here $\psi_{\rm s}=\psi(z=0)$ is the dimensionless surface potential given by
\begin{equation}
\label{surface_potential}
\psi_{\rm s} =2 {\rm sinh}^{-1} \left(\frac{2 \pi \ell_{\rm B} \sigma}{e \kappa}\right) =
-2 {\rm sinh}^{-1} \left(\frac{1}{b \kappa}\right),
\end{equation}
where $\ell_{\rm B}=e^{2}/4 \pi \varepsilon_{\rm w}k_{\rm B}T $ is the
Bjerrum length, and $b=e/2 \pi \ell_{\rm B} |\sigma|$ is the Gouy-Chapman length.

The electrostatic free energy per lipid molecule
can be calculated from the charging method~\cite{Colloid,VO}
\begin{equation}
\label{charging}
f_{\rm el}(\sigma)=\frac{k_{\rm B} T \Sigma}{e}
\int^{\sigma}_{0} {\rm d} \sigma'\, \psi_{\rm s}(\sigma'),
\end{equation}
where $\Sigma$ is the cross-sectional area per lipid.
By substituting the surface potential, Eq.~(\ref{surface_potential}), into Eq.~(\ref{charging}),
we obtain the expression
\begin{align}
\label{electro_free}
f_{\rm el} (\sigma) = \frac{k_{\rm B} T \Sigma}{\pi \ell_{\rm B} b}
\left[ b \kappa - \sqrt{1 + (b \kappa)^{2}}
 + \ln  \frac{1 + \sqrt{1+(b \kappa)^{2}}}{b \kappa}  \right] .
\end{align}
%

\begin{figure}[t!]
\begin{center}
\includegraphics[scale=0.6]{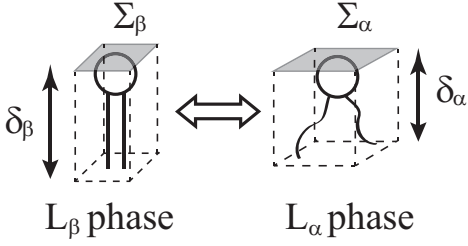}
\end{center}
\caption{Schematic illustration of molecular volume conservation.
The molecular area is indicated by a gray square,
and the molecular volume is enclosed by the dashed lines.
We assume that the molecular volume $v$ is conserved,
$v=\Sigma_{\beta} \delta_{\beta}=\Sigma_{\alpha} \delta_{\alpha}$,
for all the temperatures.
The larger the membrane thickness $\delta$, the smaller the
molecular area $\Sigma$.
}
\label{fig3}
\end{figure}

\subsection{Two-Component Charged Membranes}

In the case of membranes consisting of a binary mixture of charged
saturated (CS) and neutral unsaturated (NU) lipids,
the surface charge density, $\sigma<0$, is related to the CS molar fraction, $\phi$, by
\begin{equation}
\sigma=-\frac{e\phi}{\Sigma}.
\label{sigma}
\end{equation}
As illustrated in Fig.~\ref{fig3}, we further assume that the
lipid molecular volume, $v=\Sigma \delta$, is
conserved and stays constant at any temperature.
From Eq.~(\ref{op}), the cross-sectional area $\Sigma$ is
related to the order parameter, $m$, by
\begin{equation}
\label{coup}
\Sigma=\frac{\Sigma_{0}}{1+m},
\end{equation}
where $\Sigma_{0}$ is the cross-sectional area of a neutral
lipid in the L$_{\alpha}$ phase.
By substituting back Eqs.~(\ref{sigma}) and (\ref{coup}) into
Eq.~(\ref{electro_free}), $f_{\rm el}$ can be written as
\begin{align}
\label{electro_coup}
f_{\rm el}(m,\phi) = 2k_{\rm B}T\phi \Biggl[ \frac{1- \sqrt{1+[p \phi (1+m) ]^{2}}}{p \phi (1+m)}
\nonumber \\
+ \ln \left( p \phi (1+m) + \sqrt{1+[p \phi (1+m) ]^{2}} \right) \Biggr],
\end{align}
where $p= 2 \pi \ell_{\rm B} /\Sigma_{0}\kappa$ is a
dimensionless parameter.

The total free energy of the CS/NU membrane is finally given by
\begin{equation}
f_{\rm CS/NU}(m,\phi)=f_{\rm st}(m,\phi)+f_{\rm mix}(\phi)+f_{\rm el}(m,\phi).
\label{CS/NU-total}
\end{equation}
On the other hand, when the membrane consists of neutral saturated  (NS) and charged 
unsaturated (CU) lipids,  the total free energy of an NS/CU membrane becomes
\begin{equation}
f_{\rm NS/CU}(m,\phi)=f_{\rm st}(m,\phi)+f_{\rm mix}(\phi)+f_{\rm el}(m,1-\phi).
\label{NS/CU-total}
\end{equation}
Note that the only change between the two cases, Eqs.~(\ref{CS/NU-total}) and 
(\ref{NS/CU-total}), is that $\phi \leftrightarrow 1-\phi$ in the electrostatic free energy. 
The phase behavior of two-component charged membranes in these two cases is analyzed
using either Eq.~(\ref{CS/NU-total}) or (\ref{NS/CU-total}).

\subsection{Equilibrium Phase Behavior}

Since the electrostatic free energy described by Eq.~(\ref{electro_coup}) depends 
on the order parameter $m$, the main-transition temperature of a pure charged lipid 
(either CS or CU) is different from Eq.~(\ref{tm}).
Instead, it is given by the following conditions
\begin{align}
f_{\rm st}(m_{\beta},1)+f_{\rm el}(m_{\beta},1)
= f_{\rm st}(m_{\alpha},1)+f_{\rm el}(m_{\alpha},1),
\label{tm_charge1}
\\
\left. \frac{\partial (f_{\rm st}+f_{\rm el})}{\partial m} \right|_{m=m_{\beta}} =
\left. \frac{\partial (f_{\rm st}+f_{\rm el})}{\partial m} \right|_{m=m_{\alpha}}=0.
\label{tm_charge2}
\end{align}

For lipid mixtures, the chemical potential $\mu$ is given by
\begin{equation}
\label{chem_pote}
\left. \mu=\frac{\partial f_{\rm tot}}{\partial \phi} \right|_{m=m^{\ast}},
\end{equation}
where $f_{\rm tot}$ is taken either from Eq.~(\ref{neutral-total})
for neutral membranes or Eqs.~(\ref{CS/NU-total})--(\ref{NS/CU-total})
for the two cases of charged membranes. 
The optimal order parameter, $m^{\ast}$, is obtained by minimizing the total 
free energy with respect to $m$, $\partial f_{\rm tot}/\partial m|_{m=m^{\ast}}=0$.

\begin{table}[tbh]
\begin{center}
\caption{Parameter Values of the Model}
\begin{tabular}{ccc} 
\toprule
parameter & value & unit \\ 
\hline
$a'_{2}$ & $2.4 \times 10^{-21}$ & J/K~\cite{GL2} \\
$a_{3}$ & $-1.1 \times 10^{-18}$ & J~\cite{GL2} \\
$a_{4}$ & $2.2 \times 10^{-18}$ & J~\cite{GL2} \\
$T_{\rm S}^{\ast}$ & $-10$ & $^{\circ}{\rm C}$ \\
$T_{\rm U}^{\ast}$ & $-71$ & $^{\circ}{\rm C}$ \\
$\Sigma_{0}$ & 65 & ${\rm \AA}^2$ \\
$\varepsilon_{\rm w}$ & $7.08 \times 10^{-10}$ & ${\rm F/m}$ \\ 
\bottomrule
\label{table}
\end{tabular}
\end{center}
\end{table}

As the lipid molar fraction $\phi$ is a conserved quantity, it is convenient to 
introduce another thermodynamic potential given by
\begin{equation}
\label{thermodynamic}
g(\phi)=f_{\rm tot}(m=m^{\ast}, \phi)-\mu \phi.
\end{equation}
The two-phase coexistence region in the ($T, \phi$) plane is calculated by
using the common tangent construction to account for the constraint value
of the molar fraction $\phi$.
In other words, the two-phase coexistence has to satisfy the conditions
$g(\phi_1)=g(\phi_2)$ and $\mu(\phi_1)=\mu(\phi_2)$, where the slope of the
common tangent is equal to the chemical potential.
From the coexisting two points in the ($T, \phi$) plane, we can also construct
the phase diagram in the ($T, \mu$) plane.
The L$_{\beta}$ phase is characterized by positive $m$ values, while its
value in the L$_{\alpha}$ phase becomes negative.

\section{Results}
\label{results}

Having described our model in the previous section, we present the resulting 
phase diagrams and the temperature dependence of the molecular area.
First, we show the phase diagrams for the neutral membrane case.
This calculation essentially reproduces the theoretical results of Refs.~\cite{KSOA,KSO}.
Then, the phase diagrams for charged membranes will be presented.

\subsection{Choice of System Parameters}

As examples for typical neutral saturated  (NS) lipids and neutral unsaturated  (NU) 
lipids, we consider, respectively, dipalmitoylphosphocholine (DPPC) and 
dioleoylphosphocholine (DOPC), whose main-transition temperature is
$T_{\rm S}^{\ast\ast}=41\,^{\circ}{\rm C}$ and
$T_{\rm U}^{\ast\ast}=-20\,^{\circ}{\rm C}$, respectively.
From Eq.~(\ref{tm}), the reference temperature of NS and NU lipids
is set as  $T_{\rm S}^{\ast} = -10\,^{\circ}{\rm C}$
and $T_{\rm U}^{\ast} = -71\,^{\circ}{\rm C}$, respectively.
The molecular area of the neutral lipid in the L$_{\alpha}$ phase is fixed
to $\Sigma_{0} = 65\,{\rm \AA}^2$.
The interaction parameter is $J = 1.2 \times 10^{-20}$ or 
$1.8 \times 10^{-20}\,{\rm J}$ as considered before~\cite{KSOA,KSO,SKA1},
and the bulk salt concentration is chosen either as 
$n_{\rm b}=1\,{\rm M}$ or $10\,{\rm mM}$.
All the parameters and physical constants used in our calculations are summarized in
Table~\ref{table}.

\subsection{Two-Component Neutral Membranes}

\begin{figure*}[tbh]
\begin{center}
\includegraphics[scale=0.8]{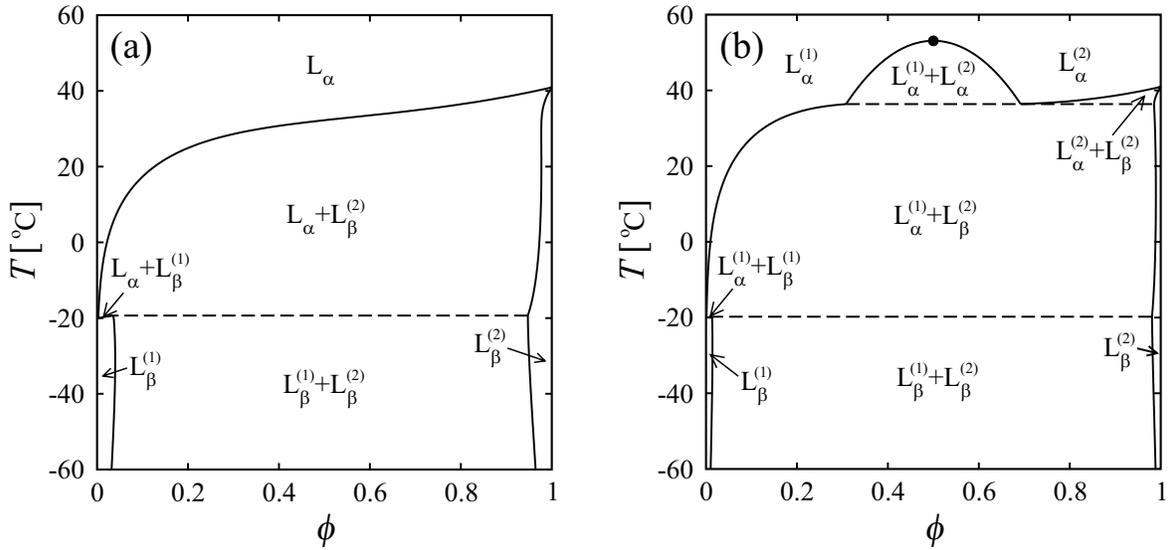}
\end{center}
\caption{Phase diagrams of a neutral NS/NU membrane as a function of the
molar fraction, $\phi$, of the NS lipids, and temperature $T$, reproducing 
the theoretical results of Refs.~\cite{KSOA,KSO}.
The attractive interaction parameter $J$ is chosen to be
$1.2 \times 10^{-20}\,{\rm J}$ in part a and $1.8 \times 10^{-20}\,{\rm J}$ in part b.
The other parameters are listed in Table~\ref{table} and are mentioned in the text.
The critical point located at ($\phi_{\rm c},T_{\rm c}$)=($0.5$, $53\,^{\circ}{\rm C}$)
in part b is indicated by a filled circle.
The horizontal dashed lines at $T=-19.2\,^{\circ}{\rm C}$ in part a, and
$T=-19.7\,^{\circ}{\rm C}$ and $T=36.4\,^{\circ}{\rm C}$ in part b indicate
a three-phase coexistence (triple point).
}
\label{fig4}
\end{figure*}

\begin{figure*}[tbh]
\begin{center}
\includegraphics[scale=0.8]{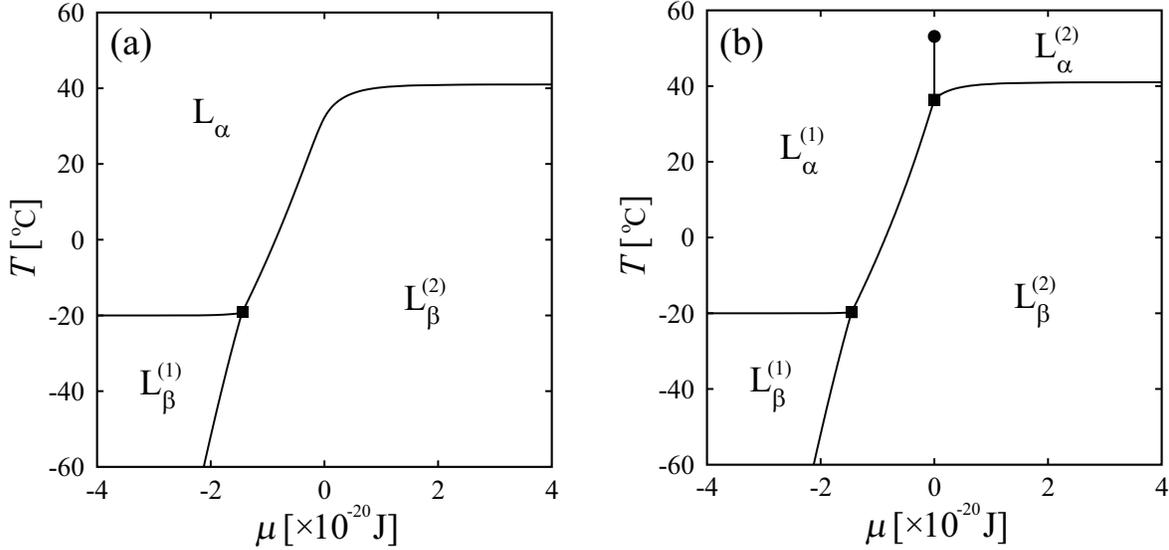}
\end{center}
\caption{Phase diagrams of a neutral NS/NU membrane as a function of the
chemical potential, $\mu$, of the NS lipids and temperature $T$.
The attractive interaction parameter $J$ is chosen to be
$1.2 \times 10^{-20}\,{\rm J}$ in part a and $1.8 \times 10^{-20}\,{\rm J}$ in part b.
The other parameters are listed in Table~\ref{table} and mentioned in the text.
The critical point located at ($\mu_{\rm c},T_{\rm c}$)=($0\,{\rm J}$, $53\,^{\circ}{\rm C}$) in part b 
is indicated by a filled circle. 
The filled squares located at ($\mu,T$)=($-1.44 \times 10^{-20}\,{\rm J}$,
$-19.2\,^{\circ}{\rm C}$) in part a, and ($-1.45 \times 10^{-20}\,{\rm J}$, $-19.7\,^{\circ}{\rm C}$) 
and ($0\,{\rm J}$, $36.4\,^{\circ}{\rm C}$) in part b indicate the triple points.
}
\label{fig5}
\end{figure*}

For the NS/NU membrane, the total free energy is given in Eq.~(\ref{neutral-total}).
As mentioned before, the NS lipid (whose molar fraction is $\phi$) has a higher
main-transition temperature than that of the NU lipid.
Following Refs.~\cite{KSOA,KSO}, we reproduce in Figs.~\ref{fig4} and \ref{fig5} the
phase diagrams of two-component NS/NU neutral membranes in the ($T,\phi$)  and
the ($T,\mu$) planes, respectively.
Figures.~\ref{fig4}(a) and \ref{fig5}(a) are calculated for
$J=1.2 \times 10^{-20}\,{\rm J}$,
while Figs.~\ref{fig4}(b) and \ref{fig5}(b) are obtained for
$J=1.8 \times 10^{-20}\,{\rm J}$.
The main-transition temperature of NS ($\phi=1$) and NU ($\phi=0$) lipids is
$T_{\rm S}^{\ast\ast}=41\,^{\circ}{\rm C}$ and
$T_{\rm U}^{\ast\ast}=-20\,^{\circ}{\rm C}$, respectively.
As shown in Fig.~\ref{fig4}(a), we obtain three two-phase coexistence regions denoted 
by L$_{\alpha}$+L$^{(1)}_{\beta}$, L$_{\alpha}$+L$^{(2)}_{\beta}$,
and L$^{(1)}_{\beta}$+L$^{(2)}_{\beta}$, and in Fig.~\ref{fig4}(b), there are five 
two-phase coexistence regions indicated by L$^{(1)}_{\alpha}$+L$^{(2)}_{\alpha}$,
L$^{(1)}_{\alpha}$+L$^{(1)}_{\beta}$,
L$^{(1)}_{\alpha}$+L$^{(2)}_{\beta}$, L$^{(2)}_{\alpha}$+L$^{(2)}_{\beta}$,
and L$^{(1)}_{\beta}$+L$^{(2)}_{\beta}$.
As discussed before~\cite{KSOA,KSO}, the phase-separated region is more extended 
when the interaction parameter $J$ is larger.

\subsection{Two-Component Charged Membranes}

\begin{figure*}[tbh]
\begin{center}
\includegraphics[scale=0.8]{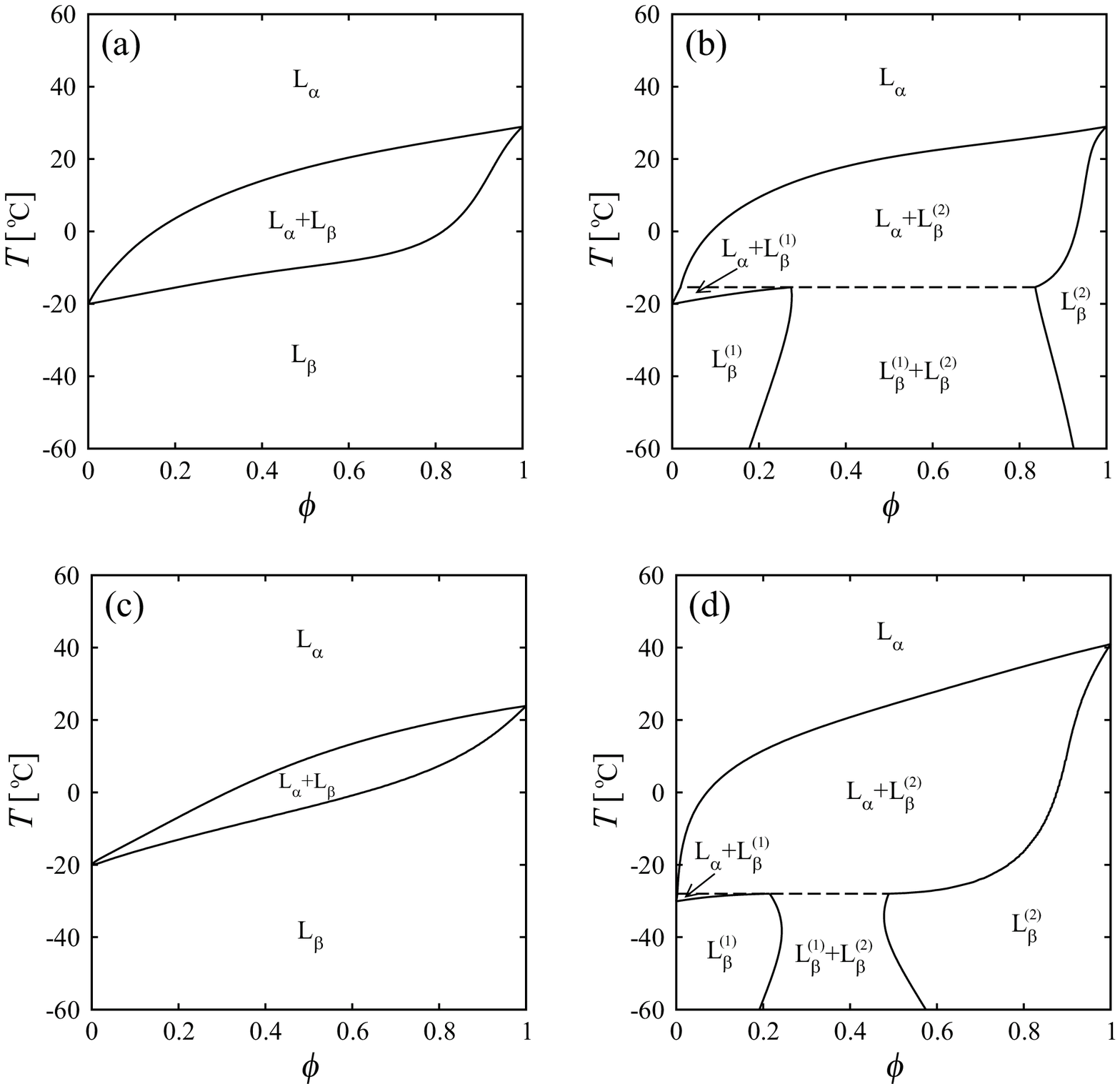}
\end{center}
\caption{Phase diagrams of two-component charged membranes as a
function of the molar fraction, $\phi$, of saturated lipid (either CS or NS)
and temperature $T$.
(a) CS/NU membrane with $J=1.2 \times 10^{-20}\,{\rm J}$ and $n_{\rm b}=1\,{\rm M}$;
(b) CS/NU membrane with $J=1.8 \times 10^{-20}\,{\rm J}$ and $n_{\rm b}=1\,{\rm M}$;
(c) CS/NU membrane with $J=1.2 \times 10^{-20}\,{\rm J}$ and $n_{\rm b}=10\,{\rm mM}$; and
(d) NS/CU membrane with $J=1.2 \times 10^{-20}\,{\rm J}$ and $n_{\rm b}=1\,{\rm M}$.
The other parameters are listed in Table~\ref{table} and are mentioned in the text.
The horizontal dashed lines located at $T=-15.3\,^{\circ}{\rm C}$ in part b and
$T=-28.0\,^{\circ}{\rm C}$  in part d indicate the three-phase coexistence (triple point).
}
\label{fig6}
\end{figure*}

\begin{figure*}[tbh]
\begin{center}
\includegraphics[scale=0.8]{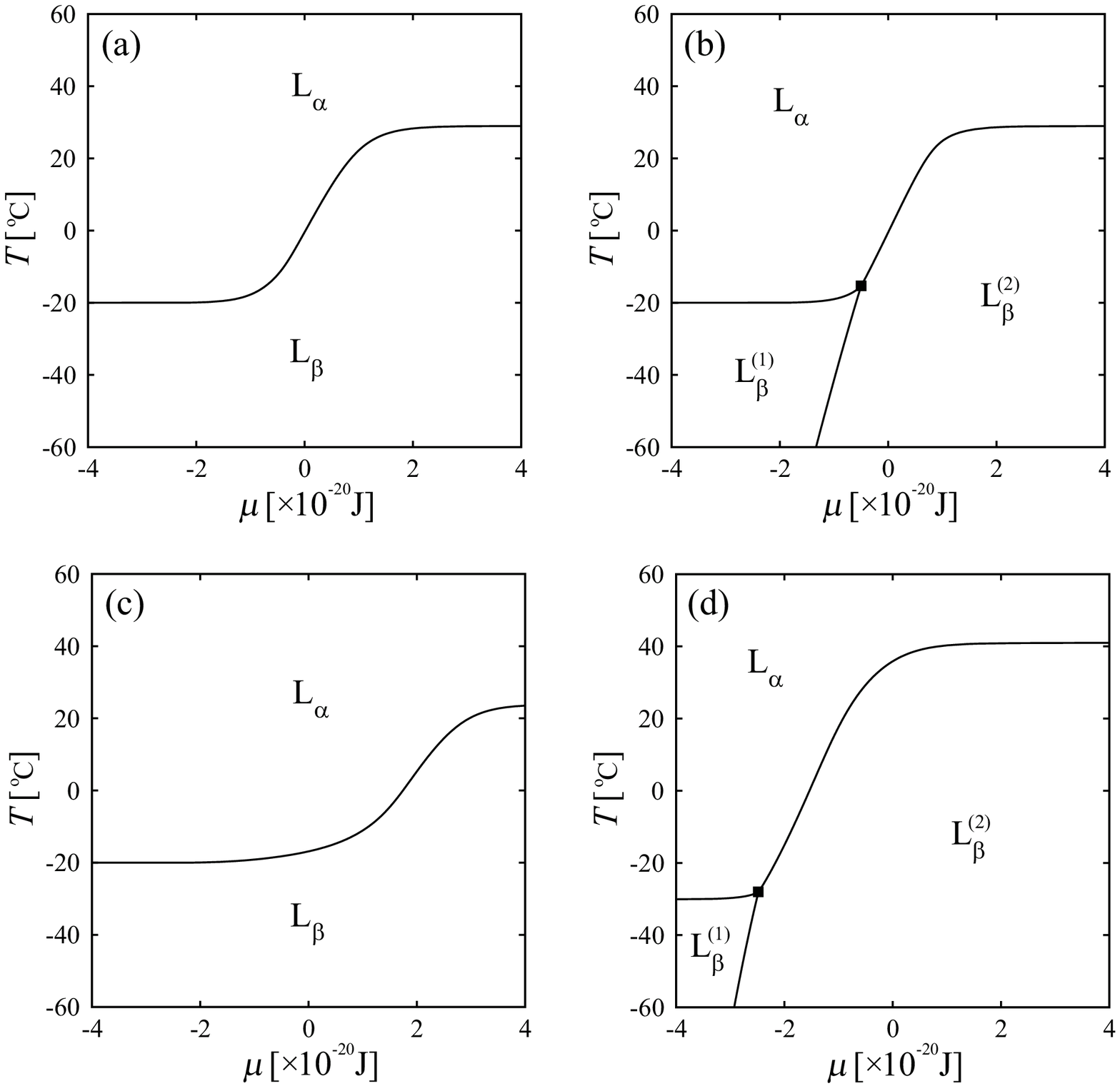}
\end{center}
\caption{Phase diagrams of two-component charged membranes as a
function of the chemical potential, $\mu$, of saturated lipid (either CS or NS)
and temperature $T$.
(a) CS/NU membrane with $J=1.2 \times 10^{-20}\,{\rm J}$ and $n_{\rm b}=1\,{\rm M}$;
(b) CS/NU membrane with $J=1.8 \times 10^{-20}\,{\rm J}$ and $n_{\rm b}=1\,{\rm M}$;
(c) CS/NU membrane with $J=1.2 \times 10^{-20}\,{\rm J}$ and $n_{\rm b}=10\,{\rm mM}$; and
(d) NS/CU membrane with $J=1.2 \times 10^{-20}\,{\rm J}$ and $n_{\rm b}=1\,{\rm M}$.
The other parameters are listed in Table~\ref{table} and are mentioned in the text.
The filled squares located at ($\mu, T$)=($-0.50 \times 10^{-20}\,{\rm J}$,
$-15.3\,^{\circ}{\rm C}$) in part b and ($-2.48 \times 10^{-20}\,{\rm J}$, $-28.0\,^{\circ}{\rm C}$) 
in part d indicate the triple points.
}
\label{fig7}
\end{figure*}

Next, we calculate the phase diagrams for two-component charged membranes.
We first consider the charged saturated/neutral unsaturated (CS/NU) (Fig.~\ref{fig1}(b)) 
lipid mixtures whose total free energy is given by Eq.~(\ref{CS/NU-total}).
In Figs.~\ref{fig6}(a) and \ref{fig7}(a), we show the calculated phase diagrams
for $J=1.2 \times 10^{-20}\,{\rm J}$ and $n_{\rm b}=1\,{\rm M}$,
recalling that $\phi$ is the molar fraction of the CS lipids.
The main-transition temperature of the CS lipid is calculated numerically
from Eqs.~(\ref{tm_charge1}) and (\ref{tm_charge2}) yielding 
$T_{\rm S}^{\ast\ast} \simeq 29\,^{\circ}{\rm C}$.
On the other hand, the main-transition temperature of the NU lipid is
$T_{\rm U}^{\ast\ast}=-20\,^{\circ}{\rm C}$, as before.
In this phase diagram, there is only one coexistence region between the
L$_{\alpha}$ and L$_{\beta}$ phases.

In Figs.~\ref{fig6}(b) and \ref{fig7}(b), we obtain the phase
diagrams for $J=1.8\times 10^{-20}\,{\rm J}$, while the other parameters
are the same as those in Figs.~\ref{fig6}(a) and \ref{fig7}(a).
The main-transition temperatures of the two lipids are unchanged, because the
conditions in Eqs.~(\ref{tm_charge1}) and (\ref{tm_charge2}) are independent
of the interaction parameter $J$.
In this case, there are three two-phase coexistence regions denoted by
L$_{\alpha}$+L$^{(1)}_{\beta}$, L$_{\alpha}$+L$^{(2)}_{\beta}$, and
L$^{(1)}_{\beta}$+L$^{(2)}_{\beta}$.

In order to further explore the effect of electrostatic interaction in CS/NU mixtures,
we looked at the case of lower salinity.
Figures~\ref{fig6}(c) and \ref{fig7}(c) are the phase diagrams corresponding
to the low salt case of $n_{\rm b}=10\,{\rm mM}$.
The other parameters are the same as those in Figs.~\ref{fig6}(a) and
\ref{fig7}(a).
The main-transition temperature of the CS lipid is now 
$T_{\rm S}^{\ast\ast} \simeq 23\,^{\circ}{\rm C}$, while
that of the NU lipid is unchanged ($T_{\rm U}^{\ast\ast}=-20\,^{\circ}{\rm C}$).
In this case, there is only one two-phase coexistence region between the
L$_{\alpha}$ and L$_{\beta}$ phases.

Finally, we show in Figs.~\ref{fig6}(d) and \ref{fig7}(d) the phase diagrams for
the neutral saturated/charged unsaturated (NS/CU) lipid mixtures whose total
free energy is given by Eq.~(\ref{NS/CU-total}).
Notice that the molar fraction $\phi$ refers here to that of the NS lipids.
The other parameters, $J$ and $n_{\rm b}$, are the same as those
in Figs.~\ref{fig6}(a) and \ref{fig7}(a).
The main-transition temperature of the NS lipid is
$T_{\rm S}^{\ast\ast}=41\,^{\circ}{\rm C}$, and that of the CU lipid is
decreased to $T_{\rm U}^{\ast\ast} \simeq -30\,^{\circ}{\rm C}$.
There are three two-phase coexistence regions denoted by
L$_{\alpha}$+L$^{(1)}_{\beta}$, L$_{\alpha}$+L$^{(2)}_{\beta}$, and
L$^{(1)}_{\beta}$+L$^{(2)}_{\beta}$.

The phase behavior of two-component membranes depends on various quantities such as
the interaction parameter $J$, the salt concentration $n_{\rm b}$, and whether
the lipid is charged or neutral.
Due to the lipid volume conservation, the main-transition temperature of the
charged lipid (either CS or CU) is reduced as compared with that of the neutral
counterpart.
These findings will be further discussed in Sec.~\ref{discussion}.

\subsection{Molecular Area in Charged Membranes}

\begin{figure*}[t!]
\begin{center}
\includegraphics[scale=0.55]{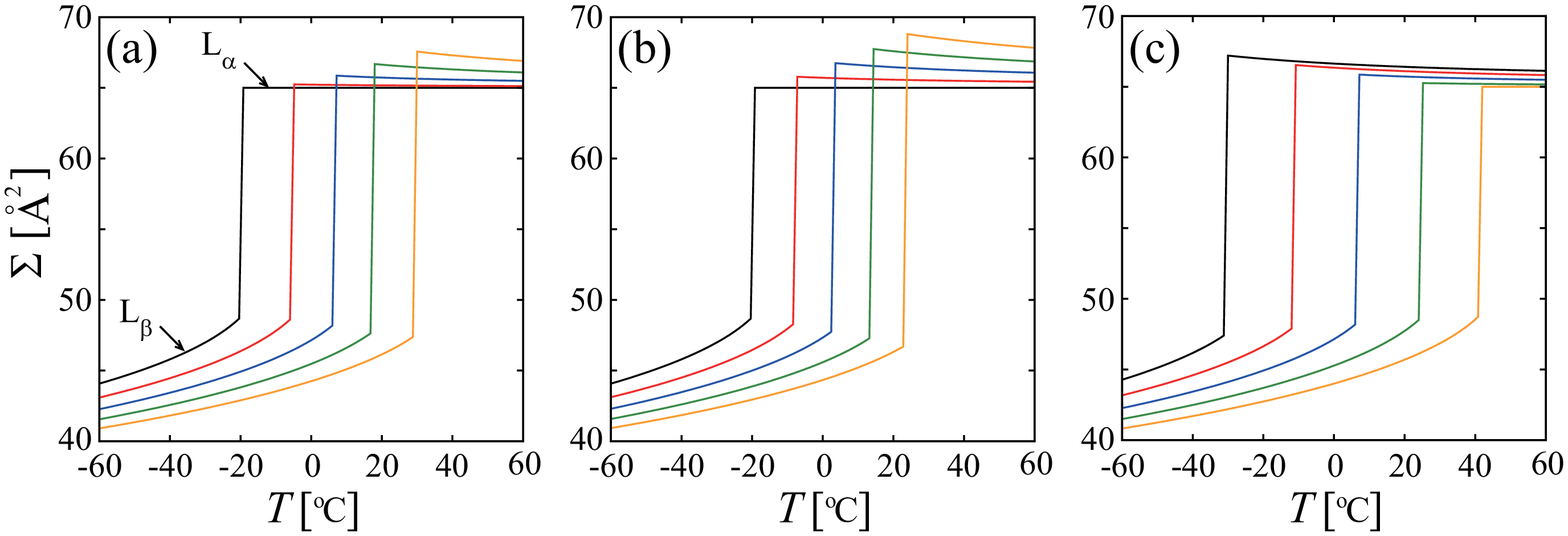}
\end{center}
\caption{Plots of molecular area $\Sigma$ as a function
of the temperature $T$.
(a) CS/NU lipid membrane with $n_{\rm b}=1\,{\rm M}$,
(b) CS/NU lipid membrane with $n_{\rm b}=10\,{\rm mM}$, and
(c) NS/CU lipid membrane with $n_{\rm b}=1\,{\rm M}$.
Black, red, blue, green, and orange lines correspond, respectively, to
$\phi=0$, $0.25$, $0.5$, $0.75$, and $1$.
}
\label{fig8}
\end{figure*}

In Fig.~\ref{fig8}, we show the temperature dependence of the molecular area,
$\Sigma$, which is calculated from the optimal order parameter, $m^{\ast}$,
and Eq.~(\ref{coup}).
For the CS/NU mixture, the molar fraction $\phi$ of the CS lipid is varied,
and the salt concentration is either $n_{\rm b}=1\,{\rm M}$ (high salt) or $10\,{\rm mM}$
(low salt), as shown in Fig.~\ref{fig8} (a) and (b), respectively.
In Fig.~\ref{fig8}(c), we show the molecular area of the NS/CU mixture
for $n_{\rm b}=1\,{\rm M}$.
Since $\Sigma$ is independent of the interaction parameter $J$, the calculation
in Fig.~\ref{fig8}(a) has been used in generating Fig.~\ref{fig6}(a) and (b).
Similarly, the calculated $\Sigma$ values shown in Fig.~\ref{fig8}(b) and (c) were used
to obtain Fig.~\ref{fig6}(c) and (d), respectively.
In order to see the change of the molecular area clearly, we ignore here the
multiphase coexistence and look at only the minimum of the total
free energy, $f_{\rm tot}$.

In Fig.~\ref{fig8}(a) and (b), for a pure NU membrane ($\phi=0$),
$\Sigma$ gradually increases when the temperature is
increased approaching the main-transition temperature.
As mentioned in Sec.~\ref{model}, the molecular area discontinuously
jumps to a larger value as appropriate for the L$_{\alpha}$ phase at
$T_{\rm U}^{\ast\ast}=-20\,^{\circ}{\rm C}$
because the main-transition is first-order.
In the L$_{\alpha}$ phase of the NU membrane, $\Sigma$ takes a
constant value, $\Sigma_{0}=65\,{\rm \AA}^2$, because the
order parameter $m^{\ast}$ always vanishes there.

In the presence of the CS lipid ($\phi>0$), $\Sigma$ also increases
and exhibits a discontinuous jump at the main-transition temperature.
At the transition temperature, as compared with the neutral membranes,
$\Sigma$ for the charged membranes is slightly smaller
for the L$_{\beta}$ phase but larger for the L$_{\alpha}$ phase.
This tendency can be clearly seen for higher molar fractions of
CS lipid ($\phi \approx 1$) or for lower salt concentrations (Fig.~\ref{fig8}(b)).
As the temperature is further increased from the main-transition temperature,
$T^{\ast\ast}$, $\Sigma$  in the L$_{\alpha}$ phase asymptotically approaches 
the constant value, $\Sigma_{0}=65\,{\rm \AA}^2$.

The temperature dependence of $\Sigma$ for the NS/CU mixture is shown in
Fig.~\ref{fig8}(c).
The molecular area first increases in the L$_{\beta}$ phase and then jumps
discontinuously at the main-transition temperature.
This molecular area expansion is larger for higher molar fraction of the CU lipid
($\phi \approx 0$).
In the next section, we shall discuss in more detail the effect of the lipid volume
conservation on the molecular area.

\section{Discussion}
\label{discussion}

Several points merit further discussion.
First, we discuss the effect of the interaction parameter $J$ on the phase behavior.
As shown in Figs.~\ref{fig4} and \ref{fig6} for neutral and charged membranes,
respectively, the phase-separated region becomes larger as $J$
increases.
In particular, we find different coexistence regions such as liquid-liquid
(L$^{(1)}_{\alpha}$+L$^{(2)}_{\alpha}$) in Fig.~\ref{fig4}(b),
and solid-solid (L$^{(1)}_{\beta}$+L$^{(2)}_{\beta}$) in
Fig.~\ref{fig6}(b).
These coexistence regions do not appear for smaller $J$ value as in
Fig.~\ref{fig4}(a) and Fig.~\ref{fig6}(a).
Such a dependence on $J$ was also reported in previous theoretical works~\cite{KSOA,KSO}.

As seen in  Fig.~\ref{fig6}, the main-transition temperature, $T^{\ast\ast}$, is reduced 
for a single-component charged lipid membrane.
The transition temperature of NS lipid is $T_{\rm S}^{\ast\ast}=  41\,^{\circ}{\rm C}$,
while it is decreased to $T_{\rm S}^{\ast\ast} \simeq 29\,^{\circ}{\rm C}$
for CS lipid as in Fig.~\ref{fig6}(a) and (b), or to $T_{\rm S}^{\ast\ast}
\simeq 23\,^{\circ}{\rm C}$ as in Fig.~\ref{fig6}(c).
The electrostatic repulsion between the charged lipid molecules tends to expand
the molecular area, resulting in a decrease of the membrane thickness, while
keeping the molecular volume $v=\Sigma \delta $ constant €€(see Fig.~\ref{fig3}).
Hence, the molecular volume conservation enhances the phase transition toward the
L$_{\alpha}$ phase, and the main-transition temperature is decreased~\cite{Jahnig}.
For a single-component unsaturated lipid, the reduction of the transition
temperature from $T_{\rm U}^{\ast\ast}=-20\,^{\circ}{\rm C}$ (for NU) to 
$-30\,^{\circ}{\rm C}$ (for CU) 	
can also be explained by the volume conservation constraint (see Fig.~\ref{fig6}(d)).

Next, we comment on the effect of the electrostatic interaction on the extension of
the phase-separated regions.
We have presented the phase behavior of a neutral NS/NU membrane (Fig.~\ref{fig4}(a)),
a charged CS/NU membrane with a high salt concentration (Fig.~\ref{fig6}(a)), and a charged
CS/NU membrane with a low salt concentration (Fig.~\ref{fig6}(c)).
The phase separation takes place below $T=41\,^{\circ}{\rm C}$ in
Fig.~\ref{fig4}(a), between $T=29\,^{\circ}{\rm C}$
and $-20\,^{\circ}{\rm C}$ in Fig.~\ref{fig6}(a), and between
$T=23\,^{\circ}{\rm C}$ and $-20\,^{\circ}{\rm C}$  in
Fig.~\ref{fig6}(c).
As the electrostatic interaction strength
is increased in the order of Figs.~\ref{fig4}(a), \ref{fig6}(a) and \ref{fig6}(c), the 
phase-separated region becomes narrower.
For the NS/CU membrane in Fig.~\ref{fig6}(d), the phase coexistence occurs
below $T=41\,^{\circ}{\rm C}$.
Although this temperature range coincides with that of neutral NS/NU membranes as
in Fig.~\ref{fig4}(a), the phase-coexistence region in charged membranes
(Fig.~\ref{fig6}(d)) is narrower than that in neutral membranes
(Fig.~\ref{fig4}(a)).
Hence, the phase separation is generally suppressed in charged membranes
as compared with neutral ones.

Some experiments~~\cite{Dimova1,SHSY,Keller4,Dimova2,Himeno} reported that
the two-phase coexistence region for charged membranes (containing mainly CU lipid)
is narrower than that for neutral membranes, and the phase separation in these
membranes can be enhanced again by adding a salt.
Compared with two-component neutral membranes (Fig.~\ref{fig4}(a)),
the phase-separated region of an NS/CU membrane (Fig.~\ref{fig6}(d))
becomes narrower due to the electrostatic interactions.
This is indeed consistent with experiments.
Previous theoretical models for charged membranes~\cite{SHSY,May,SKA2} have
considered only the contributions from the phase separation and the electrostatic
interactions.
Although these works also reported the suppression of the phase separation
in charged membranes, the main-transition was not taken into account.
In the present work, we have included this phase transition in order to fully
describe the phase behavior of two-component charged membranes.
The advantage of our model is that it describes the different phase
behavior, distinguishing CS/NU and NS/CU lipid mixtures.

As mentioned in Sec.~\ref{intro}, there are two types of two-component 
charged membranes: CS/NU and NS/CU lipid mixtures.
These two cases are presented in Fig.~\ref{fig6}(a) and (d), which manifest
that the phase-separated region in the former case is narrower than that in
the latter one.
This is because the difference in the main-transition temperature between
the saturated and unsaturated lipids is larger in the NS/CU case
(Fig.~\ref{fig6}(d)).
We note again that the reduction of the transition temperature
for charged lipids is induced by the molecular volume conservation.

In general, the larger the difference in the transition temperatures, the
stronger the segregation, and the phase separation is enhanced.
Himeno \textit {et al.}~\cite{Himeno} reported that the phase behaviors of
CS/NU and NS/CU lipid mixtures are quite different.
In particular, they obtained the liquidus lines and compared the phase
separations in CS/NU and NS/CU mixtures with that in an NS/NU mixture.
According to their results, the phase separation is enhanced for a CS/NU mixture,
whereas it is suppressed for an NS/CU mixture.
Although it is difficult to compare directly this experiment with our theory, our
model clearly distinguishes between the two different types of charged membranes,
and the trends agree with experiments.
In the previous experimental work~\cite{Himeno}, it was considered that there is 
a counterion mediated attraction between charged saturated lipids, which 
overcomes the electrostatic repulsion between the charged head groups.
For simplicity, this effect was not considered in the present study.
However, it will be of interest to consider such added interactions in future theoretical 
studies in order to make a direct comparison with the experimental results.

We further comment now on the discontinuous jump of the molecular area at the
 main-transition temperature.
For a pure NS lipid membrane ($\phi=1$ in Fig.~\ref{fig8}(c)), the molecular
area jumps from $\Sigma=48.7$ to $65\,{\rm \AA}^2$  at $T_{\rm S}^{\ast\ast}=41\,^{\circ}{\rm C}$.
For a pure CS lipid membrane ($\phi=1$ in Fig.~\ref{fig8}(a) or (b)), on the other hand,
it varies from $\Sigma=47.4$ to $67.6\,{\rm \AA}^2$ at $T_{\rm S}^{\ast\ast}=29\,^{\circ}{\rm C}$ 
in Fig.~\ref{fig8}(a), and from $\Sigma=46.9$
to $68.8\,{\rm \AA}^2$ at $T_{\rm S}^{\ast\ast}=23\,^{\circ}{\rm C}$ in Fig.~\ref{fig8}(b).
This means that the main-transition of a CS lipid occurs when the temperature is lower
and the molecular area is smaller, as compared with an NS lipid.
This behavior can be explained by the electrostatic repulsion between the charged
head groups (see Fig.~\ref{fig3}).
In general, the main-transition is induced when the attraction between the lipids
becomes weaker.
For CS lipids, the electrostatic repulsion between the lipids acts in addition to the
attraction.
Hence, the main-transition occurs at a lower temperature and smaller molecular area.
Moreover, the molecular area of a CS lipid in the L$_{\alpha}$ phase is larger than that of
an NS lipid ($\Sigma=65\,{\rm \AA}^2$) due to the electrostatic repulsion.
The increase in the molecular area at the main-transition is more enhanced when the
molar fraction of charged lipid is higher.

For higher temperatures, the molecular expansion in two-component charged membranes
is suppressed.
In the presence of charged lipid, $\Sigma$ in the L$_{\alpha}$ phase asymptotically
approaches $\Sigma_{0}$ as the temperature is further increased.
We recall here that the total free energy of CS/NU or NS/CU membranes is given by
Eq.~(\ref{CS/NU-total}) or ~(\ref{NS/CU-total}).
Taking into account the temperature dependence of
$p=2 \pi \ell_{\rm B} /\Sigma_{0}\kappa=
e/\Sigma_{0}\sqrt{8n_{\rm b}\varepsilon_{\rm w}k_{\rm B}T}$,
we note that the relative contribution of the electrostatic free energy
vanishes in the high temperature limit. 
Namely, $f_{\rm el}/(f_{\rm st}+f_{\rm mix}) \rightarrow 0$ as $T \rightarrow \infty$.
This means that the electrostatic interaction becomes irrelevant with respect
to the non electrostatic free energies in the high temperature limit,
and $\Sigma \rightarrow \Sigma_{0}$.

In neutral membranes, the optimal order parameter $m^{\ast}$
vanishes in the L$_{\alpha}$ phase.
On the other hand, in charged membranes,  $m^{\ast}$ in the
L$_{\alpha}$ phase becomes negative due to molecular volume conservation.
Since the order parameter represents the membrane thickness, negative $m$
means an expansion of the molecular area, as shown in Fig.~\ref{fig3}.
In order to understand why $m^{\ast}$ becomes negative, we expand the
electrostatic free energy $f_{\rm el}$ of Eq.~(\ref{electro_coup}) in powers of $m$
\begin{align}
\label{mex_ec}
\frac{f_{\rm el}(m,\phi)}{k_{\rm B}T} &\approx 2 \phi \left[ \frac{1-\sqrt{1+(p\phi)^{2}}}{p\phi}
+\ln \left( p \phi + \sqrt{1+(p\phi)^{2}} \right) \right] \nonumber \\
- &\frac{2\left(1-\sqrt{1+(p\phi)^{2}}\right)}{p}m
-\frac{\left(1-\sqrt{1+(p\phi)^{2}}\right)^{2}}{p\sqrt{1+(p\phi)^{2}}}m^{2}.
\end{align}
The order parameter $m$ is determined by minimizing the total free energy,
Eq.~(\ref{CS/NU-total}), which includes Eq.~(\ref{mex_ec}).
The latter equation is minimized for nonzero $m$ due to the linear term in
$m$ because $\sqrt{1+(p\phi)^{2}}>1$.
Since the stretching free energy $f_{\rm st}$ in Eq.~(\ref{stretch})  does not
include any linear term in $m$, the total free energy is minimized for a negative
$m^{\ast}<0$.
As mentioned above, the electrostatic interaction becomes negligible at higher
temperatures.
Therefore, $m^{\ast}$ asymptotically vanishes when the temperature is increased.

In this study, we have used the Poisson-Boltzmann theory to obtain the 
electrostatic potential for symmetric monovalent salt.
Such a mean-field treatment would break down when ion-ion correlations become significant. 
For example, when multivalent ions are present or a crowding state of charged lipids occurs, 
it would be important to include ion-ion correlations beyond the mean-field theory.
This point will be addressed in a future study.
Furthermore, modulated phases in neutral lipid membranes 
were discussed theoretically by taking into account both the lipid 
main transition and the spatial dependence of the lipid composition~\cite{SKA1}.
In the case of a charged lipid membrane, the formation of modulated phases was predicted within 
the Debye-H\"{u}ckel approximation without considering the lipid main transition~\cite{GA}.
However, the occurrence  of such a microphase separation has not yet been reported experimentally 
for charged bilayer membranes using optical microscopes~\cite{Dimova1,SHSY,Keller4,Dimova2,Himeno,Himeno15}.
This is because the characteristic wavelength of the microphase separation is considered to be
of the order of nanometers~\cite{Okamoto}.
Further theoretical and experimental studies are needed in order to investigate the possibility of 
such microphase separations in charged lipid membranes.

\section{Conclusions}
\label{conclusion}

In this work, we have considered phase transitions and separations in binary mixtures of
charged membranes.
Taking into account the chain main-transition, we have calculated the phase diagrams of
two-component charged and neutral lipids. 
In particular, we have considered a coupling effect between the order parameter and the
lipid composition through the molecular volume conservation.
We showed that the phase separation in charged membranes is suppressed
as compared with the neutral case.
Furthermore, as the salt concentration is increased because of screening, the phase 
separation in charged membranes is enhanced again, signaling the importance of 
electrostatic interaction.

We have focused on the combination of the lipid hydrocarbon saturation
and the charge on the headgroup (either CS/NU or NS/CU).
The phase behavior of these two types of binary mixtures of charged/neutral lipids is very
different, and is in agreement with recent experiments~\cite{SHSY,Dimova2,Himeno}.
We also find that the molecular area of charged/neutral mixtures is smaller in the L$_{\beta}$
phase and larger in the L$_{\alpha}$, as compared with neutral mixtures.
In other words, the discontinuous jump of the molecular area for charged/neutral mixtures
is larger than that for the neutral ones.
This tendency is clearly seen when the charged lipid fraction is large and/or the
salt concentration is low, and can be understood by the electrostatic repulsion between
the charged head groups.
In general, the phase behavior is significantly influenced by the presence of the
charged lipid.

Our model describes the phase behavior of binary mixtures of charged/neutral lipids.
In most of the recent experimental studies~\cite{Dimova1,SHSY,Keller4,Dimova2,Himeno},
however, the phase separation has been observed
in ternary charged lipid mixtures containing cholesterol.
Since these studies are motivated by the formation of raft domains enriched in cholesterol,
lipid mixtures containing cholesterol are also important from the biological point of view.
Putzel and Schick proposed a model~\cite{PS1} for the phase separation in neutral
membranes consisting of saturated lipids, unsaturated lipids, and cholesterol.
In the future, it will be of interest to consider a model for ternary charged lipid membranes,
which will take into account explicitly the specific role of cholesterol, in addition 
to the effect of charges in lipid mixtures.


\begin{acknowledgments}
We thank R. Okamoto for useful discussions.
N.S.\ acknowledges support from the Grant-in-Aid for Young Scientist (B) (Grant No.\ 26800222)
from the Japan Society for the Promotion of Science (JSPS) and the Grant-in-Aid for
Scientific Research on Innovative Areas ``\textit{Molecular Robotics}" (Grant No.\ 15H00806)
from the Ministry of Education, Culture, Sports, Science, and Technology of Japan (MEXT).
T.H.\ acknowledges support from the MEXT KAKENHI Grant No.\ 15H00807
and 26103516, JSPS KAKENHI Grant No.\ 15K12538, and AMED-CREST, AMED.
S.K.\ acknowledges support from the Grant-in-Aid for Scientific Research on
Innovative Areas ``\textit{Fluctuation and Structure}" (Grant No.\ 25103010) from MEXT,
the Grant-in-Aid for Scientific Research (C) (Grant No.\ 15K05250) from JSPS,
and the JSPS Core-to-Core Program ``\textit{International Research Network
for Non-equilibrium Dynamics of Soft Matter}".
D.A.\ acknowledges support from the Israel Science Foundation under Grant No.\ 438/12, 
the United States-Israel Binational Science Foundation under Grant No.\ 2012/060,
and the ISF-NSFC joint research program under Grant No.\ 885/15.
\end{acknowledgments}


\end{document}